\documentclass[aps,prb,twocolumn,superscriptaddress]{revtex4-2}

\usepackage{amsmath,amssymb,amsfonts}
\usepackage{hyperref}
\usepackage{booktabs}
\usepackage{makecell}
\usepackage{diagbox}
\usepackage{graphicx}
\usepackage{array}
\usepackage{bm}
\usepackage{physics}
\usepackage{braket}
\usepackage{subcaption}

\begin{document}

\title{Decoupling Spectral Gaps from Entanglement Complexity in Quantum Annealing}

\author{Roopayan Ghosh}
\thanks{roopayan@iitbbs.ac.in}
\affiliation{Department of Physics, School of Basic Sciences, Indian Institute of Technology Bhubaneswar, Khurda, Odisha 752051, India}
\author{Sougato Bose}
\affiliation{Department of Physics and Astronomy, University College London, Gower Street, London WC1E 6BT, England, United Kingdom}
\date{\today}

\begin{abstract}
Exponentially small spectral gaps are the central obstacle to adiabatic quantum
annealing, yet whether such bottlenecks demand correspondingly complex
entanglement remains unclear. We show that gap scaling and entanglement
complexity need not be linked. Conventional first-order transitions combine
exponentially small gaps with bounded $O(1)$ bipartite Schmidt rank, even
though the critical states retain global multipartite correlations, while
continuous transitions combine polynomially closing gaps with logarithmically
growing entanglement. We show that each pairing can also be inverted: in a known
first-order transition in a staggered-field antiferromagnetic Ising chain the gap
closes polynomially despite bounded Schmidt rank, while we construct a
\emph{mirror-twin manifold} model whose gap closes exponentially even as its
entanglement entropy grows logarithmically with system size. We derive
bounds connecting Schmidt rank to Hamming-space geometry, and identify
manifold density and connectivity as the ingredients that determine which
combination occurs. These results show that entanglement complexity alone cannot characterize
annealing difficulty, and point toward a broader search for quantum
resources that capture the geometry of state evolution in Hilbert space.
\end{abstract}
\maketitle

\section{Introduction}

Quantum annealing encodes optimization problems in the ground state of a
Hamiltonian and reaches the solution through adiabatic evolution
~\cite{10.1063/1.2995837,Matsuda_2009}. Its performance is governed by the
minimum spectral gap~\cite{Farhi2011}, which can become exponentially small
at first-order transitions~\cite{Jorg_2010}, leading to exponentially long
annealing times. A small gap makes the target state difficult to prepare,
but does this difficulty imply that the state itself must possess a
correspondingly complex quantum structure? This question has motivated the
use of quantum correlations, and in particular entanglement, as measures of
the complexity of adiabatic quantum algorithms~\cite{PhysRevA.69.052308,Hauke2015}.
The scaling of entanglement at continuous quantum phase transitions is well
established~\cite{Osterloh2002,Vidal2003}, and entanglement has been observed
and studied as a potential resource or diagnostic in quantum annealing
~\cite{PhysRevX.4.021041,RevModPhys.90.015002,capecci2026quantumresourcesnonstoquasticquantum}.
Yet it remains unclear whether the complexity of an annealing bottleneck
must be reflected in the complexity of the state traversed during the
anneal.

The central result of this work is that the scaling of the spectral gap and
the complexity of the entanglement structure are not uniquely tied to one
another. Across the annealing transitions considered here, we find that
exponentially small or polynomially closing gaps can coexist with either
bounded or logarithmically growing Schmidt structure. The four possibilities
are summarized in Table~\ref{tab:gap_entanglement}.

\begin{table}[t]
\centering
\caption{Representative examples of the four combinations of gap scaling and
entanglement complexity realized in this work.}
\label{tab:gap_entanglement}
\renewcommand{\arraystretch}{1.25}
\setlength{\tabcolsep}{3pt}

\resizebox{\columnwidth}{!}{%
\begin{tabular}{|c|c|c|}
\hline
\diagbox[width=3.0cm]{\makecell{\textbf{Gap}\\\textbf{scaling}}}
             {\makecell{\textbf{Entanglement}\\\textbf{scaling}}}
&
\makecell{\textbf{Bounded entanglement}\\[-1mm]$S=O(1)$}
&
\makecell{\textbf{Growing entanglement}\\[-1mm]$S\sim\log L$}
\\
\hline
\makecell{\textbf{Exponential gap closing}\\
$\Delta_{\min}\sim e^{-cL}$}
&
\makecell{$p$-spin;\\
structured optimization}
&
\makecell{\textbf{Mirror-twin}\\
\textbf{manifold model}}
\\[3mm]
\hline
\makecell{\textbf{Polynomial gap closing}\\
$\Delta_{\min}\sim L^{-z}$}
&
\makecell{\textbf{Staggered-field}\\
\textbf{AF Ising chain}}
&
\makecell{Continuous transition\\
($p=2$)}
\\
\hline
\end{tabular}%
}
\end{table}

This separation has a simple geometric origin. The gap is controlled by how
the annealing Hamiltonian connects competing low-energy sectors, whereas the
entanglement is controlled by how the resulting superposition resolves into
independent Schmidt branches across a bipartition. Thus, a transition may be
spectrally hard because competing sectors are separated by high-order
processes without requiring a growing Schmidt structure, while extensive
hybridization can increase the gap without producing correspondingly complex
entanglement. Conversely, a growing low-energy manifold can generate
logarithmically increasing Schmidt structure even when the relevant
inter-sector coupling remains exponentially small.

We establish this picture in several stages. In Sec.~\ref{sec:sec2}, we examine the
gap--entanglement relation in standard annealing settings, including the
$p$-spin model, non-stoquastic catalysis, and structured optimization
landscapes. These examples establish the different scaling regimes and
illustrate how changes in the connectivity of the low-energy spectrum can
modify the gap and entanglement.

In Sec.~\ref{sec:sec3}, we develop the Hilbert-space picture underlying these results.
Starting from perturbative avoided crossings, we relate the spectral gap to
Hamming-space connectivity and the entanglement to the number of independent
Schmidt branches. We then generalize this description to resonant
low-energy manifolds, derive bounds on their Schmidt structure, and use it
to understand the different cases considered in Sec.~II and to construct
the mirror-twin manifold realizing an exponentially small gap with
logarithmically growing entanglement. We further show how the same
geometric picture accounts for the polynomial gap of the staggered-field
antiferromagnetic Ising chain and the growing entanglement of continuous
transitions.

Finally, in Sec.~\ref{sec:sec4}, we discuss the implications of these
results for understanding the nature of quantum-annealing bottlenecks and
what they reveal about the quantum resources involved.
\section{Entanglement at Annealing Transitions for common models}
\label{sec:sec2}
We first examine how bipartite entanglement behaves along annealing paths with severe gap suppression. 
For
\begin{equation}
H(s)=(1-s)\sum_i\sigma_i^x+sH_P(\{\sigma_i^z\}),
\end{equation}
we track the ground state $\ket{\psi(s)}$ and its bipartite entanglement entropy
$S=-\mathrm{Tr}(\rho_A\log\rho_A)$,
for a subsystem $A$. 
We begin with the fully connected $p$-spin model, which provides a canonical setting for both first- and continuous-transition annealing, before turning to less symmetric optimization landscapes.

\begin{figure*}
    \centering

    \begin{subfigure}[b]{0.24\linewidth}
        \caption{}
        \includegraphics[width=\linewidth]{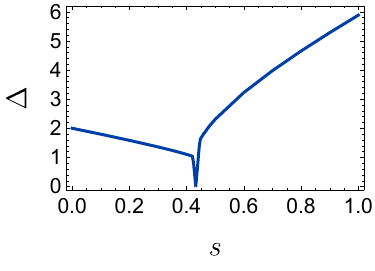}
        \label{fig:pspin_a}
    \end{subfigure}
    \hfill
    \begin{subfigure}[b]{0.24\linewidth}
        \caption{}
        \includegraphics[width=\linewidth]{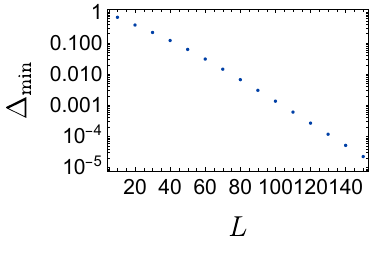}
        \label{fig:pspin_b}
    \end{subfigure}
    \hfill
    \begin{subfigure}[b]{0.24\linewidth}
        \caption{}
        \includegraphics[width=\linewidth]{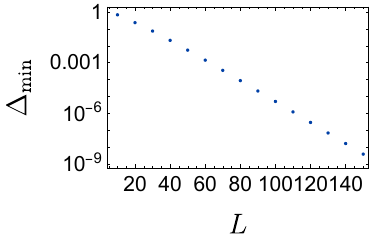}
        \label{fig:pspin_c}
    \end{subfigure}
    \hfill
    \begin{subfigure}[b]{0.24\linewidth}
        \caption{}
        \includegraphics[width=\linewidth]{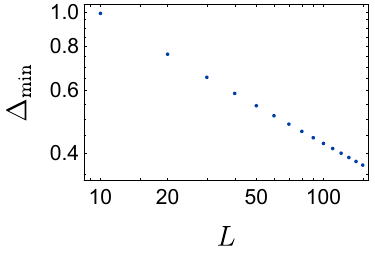}
        \label{fig:pspin_d}
    \end{subfigure}

    \vspace{2pt}

    \begin{subfigure}[b]{0.24\linewidth}
        \caption{}
        \includegraphics[width=\linewidth]{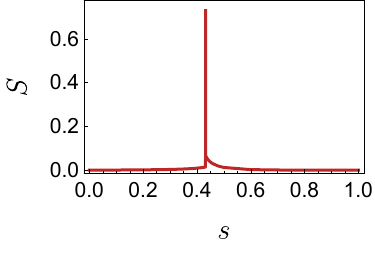}
        \label{fig:pspin_e}
    \end{subfigure}
    \hfill
    \begin{subfigure}[b]{0.24\linewidth}
        \caption{}
        \includegraphics[width=\linewidth]{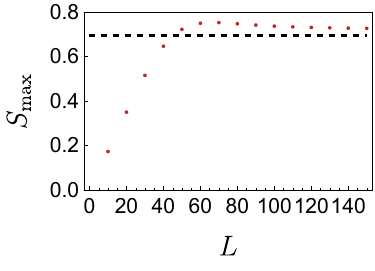}
        \label{fig:pspin_f}
    \end{subfigure}
    \hfill
    \begin{subfigure}[b]{0.24\linewidth}
        \caption{}
        \includegraphics[width=\linewidth]{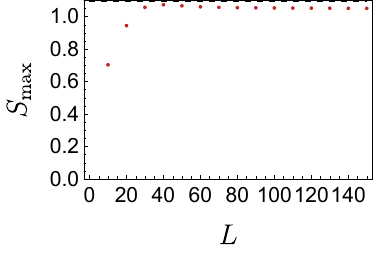}
        \label{fig:pspin_g}
    \end{subfigure}
    \hfill
    \begin{subfigure}[b]{0.24\linewidth}
        \caption{}
        \includegraphics[width=\linewidth]{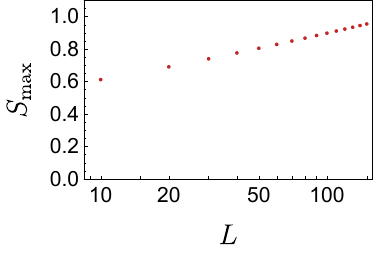}
        \label{fig:pspin_h}
    \end{subfigure}

    \caption{
    Energy gap and bipartite entanglement entropy in the fully connected $p$-spin model. 
    (a) and (e) show the energy gap $\Delta(s)$ and half-system entanglement entropy $S(s)$, respectively, as functions of the annealing parameter $s$ for $p=3$. 
    (b) and (f) display the corresponding finite-size scaling of the minimum gap $\Delta_{\rm min}$ and maximum entropy $S_{\rm max}$. 
     (c) and (g) show the same quantities for $p=4$. (d) plotted in log-log scale and (h) plotted in log-linear scale correspond to the second-order case $p=2$. 
    For the first-order transitions ($p=3$ and $p=4$), the minimum gap decreases exponentially with system size, whereas the maximum entropy approaches finite constants, $S_{\rm max}\rightarrow \ln 2$ for $p=3$ and $S_{\rm max}\rightarrow \ln 3$ for $p=4$, reflecting the coexistence of two and three competing macroscopic states at the transition, respectively. 
    In contrast, the second-order transition of the $p=2$ model exhibits polynomial gap closing, $\Delta_{\rm min}\sim L^{-z}$, and a logarithmically growing entanglement entropy, $S_{\rm max}\sim \log L$. 
    The dashed lines in (f) and (g) indicate $\ln 2$ and $\ln 3$, respectively.
    }
    \label{fig:pspin}
\end{figure*}
\subsection{The $p$-spin model}

As a paradigmatic model exhibiting both first- and second-order quantum phase
transitions, we consider the fully connected ferromagnetic $p$-spin model
~\cite{Seoane_2012_2, Nishimori2017_2, PhysRevB.95.184416, PhysRevA.98.042326},
\begin{equation}
H(s)=-(1-s)\mathcal{S}_x-s\,\frac{\mathcal{S}_z^p}{L^{p-1}},
\label{eq:pspin}
\end{equation}
where $\mathcal{S}_\alpha=\frac{1}{2}\sum_{i=1}^{L}\sigma_i^\alpha$ and
$s\in[0,1]$ is the annealing parameter. For $p\geq3$, the model undergoes a
first-order quantum phase transition, accompanied by an exponentially closing
minimum gap and hence an exponentially severe annealing bottleneck~\cite{Jorg_2010,Seoane_2012_2,Nishimori2017_2}.In contrast, $p=2$ exhibits a continuous transition with a polynomially closing
gap. The model therefore provides a natural setting in which to compare the
entanglement structure of exponentially and polynomially difficult annealing
paths. Permutation symmetry restricts the dynamics to the maximum-spin sector
$\mathcal{S}_{\rm tot}=L/2$, whose dimension grows only linearly with $L$.

We first examine the bipartite von Neumann entropy $S$ of the instantaneous
ground state for an equal bipartition, together with the minimum spectral gap.
While the dynamics can be restricted to the maximum-spin sector, evaluating
the physical bipartite entropy in this reduced representation requires a
Clebsch--Gordan decomposition, as detailed in Appendix~\ref{app:entanglement}.
Figure~\ref{fig:pspin} summarizes the results. For $p=3$, the gap develops a
sharp minimum at the first-order transition and closes exponentially with
system size [Figs.~\ref{fig:pspin}(a) and (b)]. At the same point, the
entanglement entropy develops a pronounced peak but approaches the finite
value
$S_{\rm max}\rightarrow\ln 2$
[Fig.~\ref{fig:pspin}(f)]. Thus, the exponentially small gap does not require
a growing bipartite entanglement. The small finite-size deviations arise from
weak participation of additional states.

The same exponential gap closing occurs for $p=4$ [Fig.~\ref{fig:pspin}(c)],
but now
$S_{\rm max}\rightarrow\ln 3$
[Fig.~\ref{fig:pspin}(g)]. The additional Schmidt branch originates from the
symmetry-related degeneracy of the ferromagnetic phase, as discussed below.
Thus, despite their exponentially severe annealing bottlenecks, both
first-order cases retain an $O(1)$ entanglement structure.

\begin{figure}
    \centering

    \begin{subfigure}[b]{0.47\linewidth}
        \centering
         \caption{}
        \includegraphics[width=\linewidth]{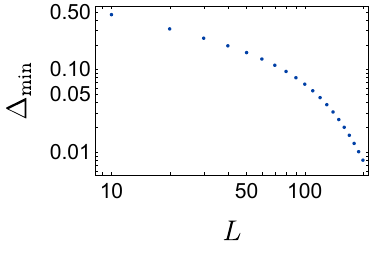}
       
        \label{fig:gap_p3k3}
    \end{subfigure}
    \hfill
    \begin{subfigure}[b]{0.47\linewidth}
        \centering
        \caption{}
        \includegraphics[width=\linewidth]{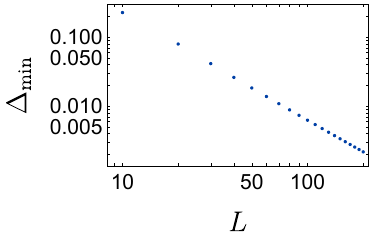}
        
        \label{fig:gap_p5k5}
    \end{subfigure}

    \vspace{0.5em}

    \begin{subfigure}[b]{0.43\linewidth}
        \centering
          \caption{} 
        \includegraphics[width=\linewidth]{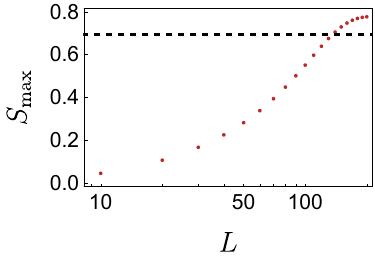}
        
        \label{fig:entropy_p3k3}
    \end{subfigure}
    \hfill
    \begin{subfigure}[b]{0.43\linewidth}
        \centering
             \caption{}
        \includegraphics[width=\linewidth]{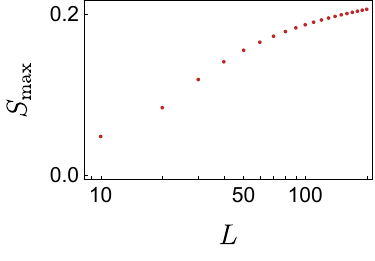}
       
        \label{fig:entropy_p5k5}
    \end{subfigure}

    \caption{
    Finite-size scaling of the minimum energy gap $\Delta_{\rm min}$ and
    maximum bipartite entanglement entropy $S_{\rm max}$ for non-stoquastically
    catalyzed fully connected $p$-spin models. The top row shows the minimum
    gap for (a) $p=3$, $k=3$ and (b) $p=5$, $k=5$, while the bottom row shows
    the corresponding maximum bipartite entanglement entropy for (c) $p=3$,
    $k=3$ and (d) $p=5$, $k=5$. The gaps are shown on a log-log scale and (d) is plotted in log-linear scale.
    The dashed line in (c) indicate the first-order value
    $S_{\rm max}=\ln 2$.
    }
    \label{fig:pspin2}
\end{figure}
For $p=2$, the minimum gap instead closes polynomially
[Fig.~\ref{fig:pspin}(d)], while
$S_{\rm max}\sim\log L$
[Fig.~\ref{fig:pspin}(h)]. The critical state therefore develops an
increasing Schmidt structure as the system grows, in contrast to the
finite-branch structure found for $p\geq3$. This logarithmic scaling is a
known property of the critical Lipkin--Meshkov--Glick model, to which the
$p=2$ Hamiltonian belongs~\cite{Ors2008}. For an equal bipartition, the
asymptotic scaling takes the form $S_{\rm max} \simeq \frac{1}{6}\ln L+\mathrm{const.}$,
although substantial finite-size corrections can persist to large system
sizes~\cite{Ors2008}. Consequently, over the system sizes accessible here,
the entropy can remain comparable in magnitude to the finite limiting
values of the first-order cases, $S_{\rm max}\to\ln 2$ and $\ln 3$ for
$p=3$ and $p=4$, respectively. The distinction is instead asymptotic:
the entropy of the $p=2$ critical state continues to grow logarithmically,
whereas the first-order cases approach finite constants. For the scaling
observed in Fig.~\ref{fig:pspin}(h), this logarithmic growth becomes
comparable to the $\ln 3$ saturation value only at system sizes of order
a few hundred, illustrating why the difference is not yet dramatic on the
scale of the figure.

The $p$-spin model thus establishes the two conventional regimes: an
exponentially closing gap with bounded Schmidt structure, and a polynomially
closing gap with growing Schmidt structure. The key question is whether this
pairing is intrinsic to the gap scaling itself, or instead reflects the
specific connectivity of the low-energy states. We next modify that
connectivity to distinguish these possibilities.

Non-stoquastic catalysts~\cite{ Seki2012,Seki_2015,Nishimori2017_2, PhysRevB.95.184416} provide a direct way to probe this question. By adding higher-order transverse interactions, they modify the connectivity of the low-energy manifold and can soften the first-order transition. We consider
\begin{equation}
H(s)=-(1-s)\mathcal{S}_x-s\frac{\mathcal{S}_z^p}{L^{p-1}}
+s(1-s)\frac{\mathcal{S}_x^k}{L^{k-1}},
\label{eq:pspincatalyst}
\end{equation}
where the last term is a non-stoquastic catalyst. 
Figure~\ref{fig:pspin2} compares the $p=3$, $k=3$ and $p=5$, $k=5$ cases.

For $p=3$, $k=3$, the catalyst substantially enhances the minimum gap, while
the maximum entanglement grows beyond the $\ln 2$ value of the uncatalyzed
first-order transition. However, the finite-size scaling does not yet exhibit
the clear polynomial gap closing and logarithmic entropy growth of the
continuous $p=2$ transition. The resulting behaviour therefore represents a
softened first-order regime in which both the spectral bottleneck and the
Schmidt structure are modified.

For $p=5$, $k=5$, the behaviour is qualitatively different. The minimum gap
is consistent with polynomial closing, while the entanglement grows
systematically with system size, as expected for a continuous transition
~\cite{Nishimori2017_2}. Thus, the catalyst can drive the system from the
finite-branch, exponentially suppressed regime towards a growing low-energy
manifold with substantially enhanced gap and entanglement.

These examples suggest that gap enhancement and entanglement growth can be
related through the connectivity of the low-energy manifold. In particular,
increasing the connectivity can simultaneously soften the spectral
bottleneck and enhance the Schmidt structure. This raises the question of whether this apparent connection is intrinsic to the gap and entanglement themselves, or instead reflects how the annealing Hamiltonian connects the relevant low-energy states — a question we answer in Sec.~\ref{sec:sec3} by showing it is the latter. We next turn to structured optimization problems,
where the competing low-energy configurations are no longer constrained by
the collective symmetry of the $p$-spin model.
\subsection{Structured models}

\begin{figure*}
    \centering
    \subfloat[]{%
        \includegraphics[width=0.235\linewidth]{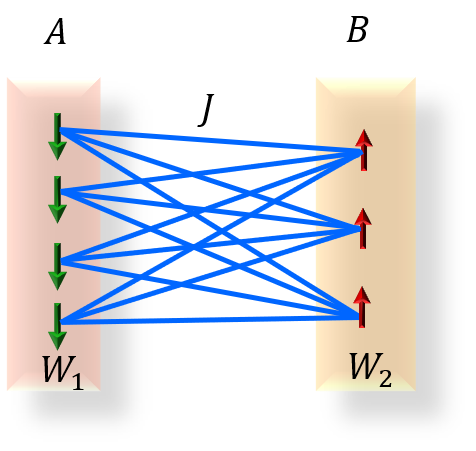}
    }\hfill
    \subfloat[]{%
        \includegraphics[width=0.235\linewidth]{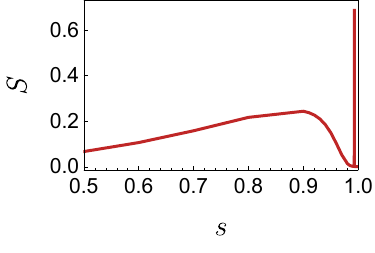}
    }\hfill
    \subfloat[]{%
        \includegraphics[width=0.235\linewidth]{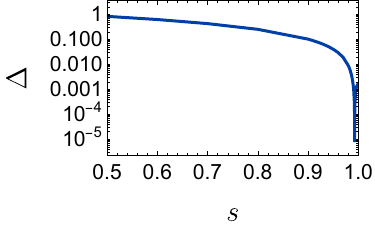}
    }\hfill
    \subfloat[]{%
        \includegraphics[width=0.235\linewidth]{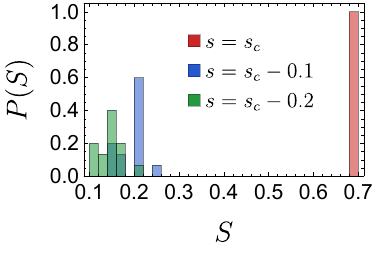}
    }
   \caption{
Entanglement structure of a representative bipartite graph instance
exhibiting a first-order quantum phase transition.
\textbf{(a)} Schematic of the bipartite graph, showing the natural
partition into subsystems $A$ and $B$, with intra-partition weights
$W_1=0.04$ and $W_2=0.0396$, and inter-partition coupling $J=0.2132$.
\textbf{(b)} Bipartite entanglement entropy across the natural $5|6$
graph partition for $L=11$ spins as a function of the annealing parameter
$s$. The entropy develops a broad enhancement at intermediate $s$,
associated with the delocalization of the annealing state, before
decreasing as the system approaches the first-order transition.
\textbf{(c)} Instantaneous energy gap $\Delta$ between the ground and
first excited states for the same $L=11$ system as a function of $s$,
showing the strongly suppressed avoided crossing associated with the
first-order transition.
\textbf{(d)} Histogram of the bipartite entanglement entropy over all
$1023$ inequivalent physical-spin bipartitions at $s=s_c$, $s_c-0.1$,
and $s_c-0.2$, showing the redistribution of entanglement across
different bipartitions as the system approaches the first-order
transition.
}
    \label{fig:bipartite}
\end{figure*}

We now turn to optimization-inspired Hamiltonians~\cite{BAPST2013127,
Pecci_2024, choi2008minorembeddingadiabaticquantumcomputation} whose
low-energy spectrum exhibits perturbative avoided crossings rather than
collective phase coexistence. As a representative example, we consider the
bipartite weighted graph introduced in Ref.~\cite{feinstein2023,Nutricati2025,
hnzj-j7kt}, originally motivated by the Maximum Weighted Independent Set
(MWIS) problem. As illustrated schematically in Fig.~\ref{fig:bipartite}(a),
the graph consists of two partitions, $A$ and $B$, connected by all-to-all
inter-partition couplings, with appropriately chosen vertex weights that
tune the system to a perturbative avoided crossing during the anneal. The
precise graph construction and Hamiltonian are given in
Appendix~\ref{app:appA}. 

Figure~\ref{fig:bipartite} illustrates the corresponding entanglement
structure. Panel~(b) shows the von Neumann entropy across the natural
bipartition $A|B$ shown in panel~(a). At $s=0$, the ground state is the
product paramagnetic state and the bipartite entropy vanishes. As the
anneal progresses, the transverse-field driver dresses the instantaneous
ground state with computational-basis configurations of the problem
Hamiltonian, producing a gradual increase of the entanglement. The entropy
develops a broad maximum at intermediate $s$ and subsequently decreases as
the ground state becomes increasingly localized in the computational basis,
vanishing again at $s=1$.

This behavior differs from the collective first-order transitions discussed
above. In the fully connected $p$-spin model, the relevant low-energy
states at the transition are well described by a small number of
macroscopically distinct branches, giving $S_{\max}\to\ln 2$ for $p=3$ and
$S_{\max}\to\ln 3$ for $p=4$. Here, by contrast, appreciable bipartite
entanglement develops well before the minimum gap is reached. The
intermediate-$s$ entropy reflects the perturbative dressing of the
computational-basis states by virtual excitations, rather than the
coexistence of a small number of competing macroscopic phases.

Close to the avoided crossing, the entropy across the natural partition
rises sharply toward $\ln 2$, consistent with a two-branch description of
the resonant low-energy sector. This occurs at the same value of $s$ at
which the instantaneous gap exhibits its pronounced minimum in
Fig.~\ref{fig:bipartite}(c). The exponentially small scaling of this
minimum gap with system size has been established for these perturbative
crossings~\cite{hnzj-j7kt}; the present figure instead illustrates the
associated entanglement structure for a representative instance. At the
crossing, the relevant low-energy sector is effectively two-dimensional,
so that the ground state acquires approximately equal weight on two
competing computational configurations, producing the finite
two-branch entanglement characteristic of a Schr\"odinger-cat-like state.

Panel~(d) examines whether this two-branch structure is specific to the
natural graph partition. For the $L=11$ instance shown here, we evaluate
the bipartite entropy for all inequivalent bipartitions of the physical
spins. There are
$2^{L-1}-1=1023$ such bipartitions, where complementary partitions
$A|B$ and $B|A$ are counted only once. At the critical point,
$s=s_c$, the resulting entropy distribution is sharply concentrated near
$\ln 2$. Thus, the finite-branch structure at the avoided crossing is not
peculiar to the natural partition displayed in panel~(a), but persists
across essentially all physical bipartitions of this instance.

Away from the avoided crossing, the distribution broadens substantially.
At $s=s_c-0.1$ and $s=s_c-0.2$, different bipartitions therefore resolve
different amounts of the perturbative dressing, producing a broad range
of entanglement entropies. Since $s_c\simeq0.99$, these values correspond
to the intermediate-$s$ regime in panel~(b). The evolution from a broad
distribution of bipartite entanglement to a sharply concentrated
$\ln 2$ distribution illustrates the crossover from a broadly dressed
ground state to a small resonant manifold at the perturbative crossing.

\begin{figure*}
    \centering
    \includegraphics[width=0.32\linewidth]{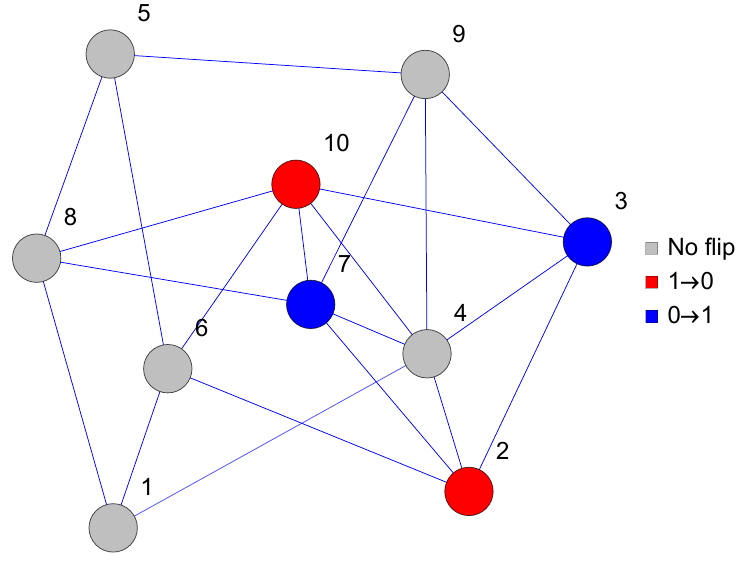}
    \includegraphics[width=0.28\linewidth]{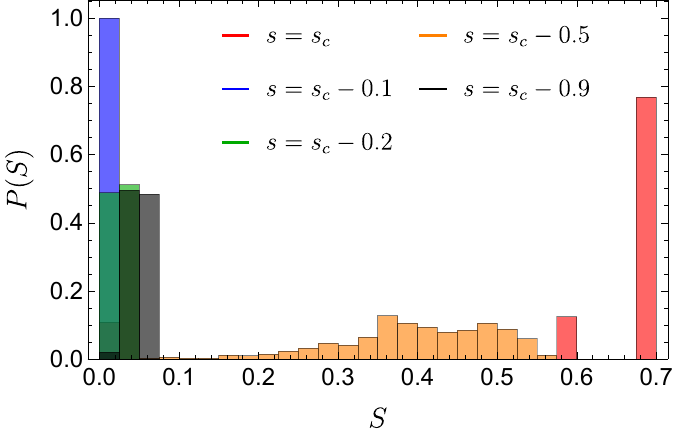}
    \includegraphics[width=0.36\linewidth]{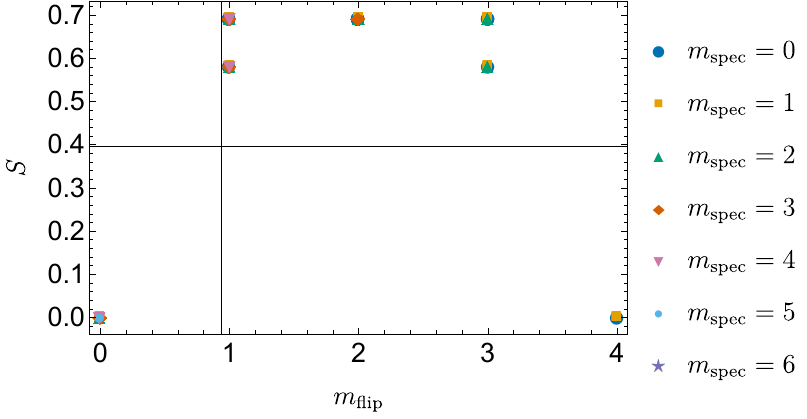}
   \caption{
Entanglement across bipartitions for a representative random graph instance at a first-order quantum phase transition.
\textbf{Left:} Graph corresponding to the avoided crossing, with vertices colored according to the two dominant computational basis configurations participating in the transition. Gray vertices remain unchanged, while red and blue denote spins that change their orientation between the two configurations.
\textbf{Center:} Histogram of the bipartite entanglement entropy over all $511$ inequivalent bipartitions at different points along the annealing path, namely $s=s_c$, $s_c-0.1$, $s_c-0.2$, $s_c-0.5$, and $s_c-0.9$.
\textbf{Right:} Bipartite entanglement entropy at the critical point $s=s_c$ for every inequivalent bipartition, plotted as a function of the number of flipped spins contained within the chosen subsystem, $m_{\mathrm{flip}}$. The color of each point indicates the corresponding number of spectator spins, $m_{\mathrm{spec}}$, contained in the same subsystem.
}
    \label{fig:random}
\end{figure*}
\subsection{Random graph}

We now consider a random graph instance, representative of a
spin-glass optimization landscape. The couplings and local fields are chosen
such that the annealing path exhibits a perturbative avoided crossing, with
the detailed graph construction given in Appendix~\ref{app:appB}. Unlike the
highly symmetric bipartite example discussed above, no symmetry relates the
competing computational-basis configurations, so different bipartitions can
resolve different aspects of the avoided crossing.

The left panel of Fig.~\ref{fig:random} shows the two dominant
computational-basis configurations at the avoided crossing,
\begin{equation}
\ket{z_2}=\ket{1100100001},
\qquad
\ket{z_3}=\ket{1010101000},
\end{equation}
which form the dominant two-state component and differ on four spins,
$d_H(z_2,z_3)=4$. The red and blue vertices identify these four active
spins, while the remaining vertices are spectators. At the crossing, the
corresponding amplitudes are approximately $0.625$ and $0.707$, accounting
for about $39\%$ and $50\%$ of the wavefunction weight, respectively. The
exact ground state also contains a third configuration,
$\ket{z_1}=\ket{1100100000}$, with amplitude $0.328$ and hence about
$11\%$ weight. This state satisfies
$d_H(z_1,z_2)=1$ and $d_H(z_1,z_3)=3$, providing a controlled departure
from the ideal two-state picture.

The middle panel shows the distribution of bipartite entanglement over all
inequivalent bipartitions. At the avoided crossing, the distribution remains
concentrated near $S=\log 2$, but acquires a finite width compared with the
highly symmetric bipartite example. Thus, although three computational-basis
configurations have appreciable weight, the entanglement does not develop a
corresponding $\log 3$ plateau.

The right panel resolves this structure by plotting the entropy against the
number of active spins, $m_{\rm flip}$, contained in the chosen subsystem.
Spectator spins contribute only product factors and therefore do not alter
the Schmidt spectrum. Consequently, cuts with the same $m_{\rm flip}$ but
different numbers of spectator spins give essentially identical entropies.
For the dominant two-state component, $m_{\rm flip}=0$ or $4$ places all
active spins on one side of the cut and therefore gives vanishing
entanglement, whereas $1\leq m_{\rm flip}\leq3$ separates the two dominant
configurations and produces the $\log 2$-like entanglement values observed
in the plot.

The third configuration does not generate an additional independent
Schmidt branch. Since $\ket{z_1}$ and $\ket{z_2}$ differ by only one spin,
their superposition is separable across every bipartition: the differing
spin lies entirely on one side of the cut and can be absorbed into a local
state. Thus, the $\ket{z_1}$ component modifies the Schmidt coefficient
associated with the $\ket{z_2}$ branch rather than introducing a new branch.
Its unequal weight and asymmetric position in Hamming space then produce
the observed cut-dependent deviations from $\log 2$, without generating a
$\log 3$ plateau.

This random graph therefore provides a concrete example in which several
computational-basis configurations participate appreciably in an avoided
crossing without producing the same number of independent Schmidt branches.
The relevant quantity for the entanglement is not simply the number of
participating configurations, but how their Hamming-space geometry projects
onto a given bipartition. This distinction motivates the general manifold
description developed below.
\section{Gap Scaling and Hilbert-Space Geometry}
\label{sec:sec3}

The examples above suggest that the relation between spectral gap and
entanglement is tied to how the annealing Hamiltonian connects the relevant
low-energy states. However, the nature of this relation is not yet clear.
In particular, the same modification of the low-energy connectivity can
affect both the gap and the entanglement, while the random-graph example
shows that several configurations can participate in an avoided crossing
without producing a correspondingly larger Schmidt structure. We therefore
turn to the underlying Hilbert-space geometry to determine which features
of the low-energy states control the two quantities.

We begin with perturbative avoided crossings, where the relevant
low-energy configurations are connected by virtual spin-flip processes.
This provides a simple setting in which the relation between their
separation in Hilbert space, the resulting spectral gap, and the
entanglement of the hybridized ground state can be examined explicitly.
We then extend this picture to more general low-energy structures, including
resonant manifolds and continuous transitions, to determine how the
different combinations of gap and entanglement identified above can arise.

\subsection{Two-state avoided crossings and perturbative tunneling}
\label{sec:twostate}

In the problem-dominated regime, low-energy eigenstates are perturbatively
connected to computational-basis configurations,
\begin{equation}
\ket{z}=\ket{z_1z_2\cdots z_L},
\qquad
z_i\in\{\uparrow,\downarrow\}.
\end{equation}
For a transverse-field driver
$H_D=-(1-s)\sum_i\sigma_i^x$,
a single application flips one spin, so two configurations separated by
Hamming distance
$d=d_H(z_1,z_2)$
first couple at order $d$ in perturbation theory. The effective tunneling
amplitude is
\begin{equation}
t_{\rm eff}
\sim
\sum_{\mathcal P}
\prod_{m=1}^{d}
\frac{1-s}{\Delta E_m},
\label{eq:teff}
\end{equation}
where $\mathcal P$ denotes virtual spin-flip paths. For finite intermediate
energy denominators this gives, up to polynomial path-counting factors,
\begin{equation}
t_{\rm eff}\sim (1-s)^d,
\qquad
\Delta_{\min}\sim 2|t_{\rm eff}|.
\label{eq:gappert}
\end{equation}
Since $s \sim 1$ in this regime, a Hamming distance $d$ that grows
extensively with system size leads to an exponentially small tunneling
amplitude when the intermediate energy denominators remain $O(1)$. This is
the mechanism underlying the perturbative avoided crossings in both the
structured bipartite and random-graph examples discussed above.

At the center of the crossing,
\begin{equation}
\ket{\psi}
\simeq
\frac{1}{\sqrt2}
\left(
\ket{z_1}+\ket{z_2}
\right),
\end{equation}
so that
\begin{equation}
r_S\leq2,
\qquad
S\leq\log2,
\end{equation}
with the bound saturated whenever the two configurations differ across the
bipartition.

The Hamming distance therefore determines the perturbative order at which
the gap opens, but does not fix the entanglement. For example, the states
\begin{equation}
\ket{z_1}
=
\ket{\uparrow\uparrow\uparrow\uparrow\uparrow\uparrow},
\qquad
\ket{z_2}
=
\ket{\downarrow\downarrow\downarrow\uparrow\uparrow\uparrow},
\end{equation}
are separated by $d=L/2$ and can give $S=\log2$ across any cut that
intersects the active region. The maximally separated pair
\begin{equation}
\ket{\psi}
\simeq
\frac{1}{\sqrt2}
\left(
\ket{z}+\ket{\bar z}
\right),
\qquad d=L,
\end{equation}
has the strongest perturbative suppression for a transverse-field driver,
but the same bipartite entropy. Moreover, $d=L$ is special: every
configuration has a unique partner at this distance, namely its complement.
Thus, the maximally separated pair does not require a larger Schmidt
structure despite giving the highest perturbative order.

The geometry of the entanglement distinguishes the spatial structure of
these crossings. For a globally rearranged pair such as
\begin{equation}
\ket{\psi}
\simeq
\frac{1}{\sqrt2}
\left(
\ket{\uparrow}^{\otimes L}
+
\ket{\downarrow}^{\otimes L}
\right),
\label{eq:globalcat}
\end{equation}
every nontrivial bipartition has $S_A=\log2$. By contrast,
\begin{equation}
\ket{\psi}
\simeq
\frac{1}{\sqrt2}
\left(
\ket{\uparrow}^{\otimes L}
+
\ket{\downarrow\downarrow\downarrow\uparrow\uparrow\uparrow}
\right)
\label{eq:localcat}
\end{equation}
factorizes into an entangled active region and spectator spins, so only
bipartitions intersecting the active region carry entropy. The same peak
entropy can therefore correspond to very different rearrangements in
configuration space.

Additional computational branches need not increase the Schmidt dimension.
For example, across a bipartition $A|B$,
\begin{equation}
\ket{\psi}
=
a\ket{0_A0_B}
+b\ket{1_A1_B}
+c\ket{1_A0_B}
\end{equation}
has a coefficient matrix of rank two for $a,b\neq0$, and hence
$r_S=2$ and $S_A\leq\log2$. The additional configuration redistributes
the two Schmidt weights rather than creating a new Schmidt sector.
Different bipartitions can therefore acquire different entropies, including
values below $\log2$, without producing a $\log3$ plateau. This is precisely
the structure observed in the random-graph example, where a third
computational configuration has appreciable weight but does not generate an
independent third Schmidt branch.

The few-state examples therefore show that neither Hamming distance nor the
number of populated computational configurations alone determines the
entanglement. What matters is how the participating configurations resolve
into independent Schmidt branches across a given cut. We now ask how this
picture changes when several low-energy configurations become resonantly
connected and form a genuine low-energy manifold.

\begin{figure}
    \centering
    \subfloat[]{%
        \includegraphics[width=0.45\linewidth]{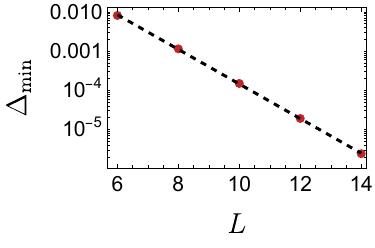}
    }\hfill
    \subfloat[]{%
        \includegraphics[width=0.45\linewidth]{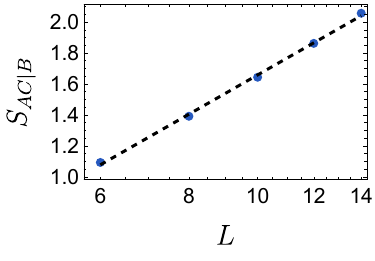}
    }
    \caption{
    Scaling of the gap at avoided level crossing and bipartite entanglement for the constructed
    resonant-manifold annealing Hamiltonian. The numerical results are
    obtained for $\Gamma=1$, $K=10$, $U_G=30$, $U_E=0.5$, and
    $\delta=0.03n$, with $n=3,\ldots,6$ and $L=2n$, alongside the one selector spin making system sizes $2 n+1$ being simulated. the crossing occurs at $s_c\sim 0.57-0.61$, drifting towards the larger value with increasing system sizes.
    (a) The minimum gap decreases exponentially with system size, with the
    dashed line showing a fit to
    $\ln\Delta_{\min}=1.316-1.017L$, corresponding to
    $\Delta_{\min}\sim e^{-1.017L}$.
    (b) The entanglement entropy across the $(A,C)|B$ bipartition grows
    logarithmically with system size, with the dashed line showing a fit to
    $S_{AC|B}=-0.958+1.136\ln L$.
    Thus, the construction realizes a growing Schmidt structure,
    $S_{AC|B}\sim\ln L$, while the minimum gap is exponentially suppressed.
    }
    \label{fig:designed}
\end{figure}

\subsection{From dilute manifolds to the \emph{mirror-twin manifold}}
\label{sec:mirrortwin}
The perturbative picture developed above corresponds to a \emph{dilute}
low-energy manifold, in which only a small number of computational-basis
configurations participate in the relevant avoided crossing. We call a manifold dilute \emph{dilute} when its number
of configurations remains $O(1)$ as $L$ grows~\cite{Imbrie2016}. More generally,
however, an optimization problem can possess a degenerate or nearly
degenerate manifold containing an increasing number of low-energy
configurations.  For the two
competing manifolds introduced below, we quantify their size through the
densities
\begin{equation}
\eta_G \equiv \frac{R_G}{2^L}, \qquad
\eta_E \equiv \frac{R_E}{2^L},
\label{eq:resdensity}
\end{equation}
where $R_G$ and $R_E$ denote the numbers of computational configurations
in the respective manifolds. Thus, dilute manifolds have
$\eta_G,\eta_E\sim2^{-L}$, whereas the regime of interest here is one in
which $R_G$ and/or $R_E$ grows with system size.

To make this generalization explicit, consider two sets of low-energy
configurations,
\begin{equation}
\mathcal G=\{\ket{g_\mu}\}_{\mu=1}^{R_G},
\qquad
\mathcal E=\{\ket{e_\nu}\}_{\nu=1}^{R_E},
\end{equation}
described near an avoided crossing by
\begin{equation}
H_{\rm eff}=
\begin{pmatrix}
H_{\mathcal G} & T\\
T^\dagger & H_{\mathcal E}
\end{pmatrix},
\label{eq:Heffmanifold}
\end{equation}
where, for a transverse-field driver,
$T_{\mu\nu}\sim(1-s)^{d_H(g_\mu,e_\nu)}$ up to energy-denominator and
path-counting factors. Thus, the coupling between the competing manifolds
can remain exponentially small even when states within each manifold are
connected by much lower-order processes.

A state supported on many computational configurations does not, however,
necessarily have a large Schmidt rank. Writing
\begin{equation}
\ket{\psi}
=
\sum_\mu a_\mu\ket{g_\mu}
+
\sum_\nu b_\nu\ket{e_\nu},
\end{equation}
gives the general bound\footnote{Note that this bound is consistent with the coexistence of two phases at the transition point in a typical first order phase transition~\cite{Continentino2004}.}
\begin{equation}
r_S\le R_G+R_E,
\qquad
S\le\log(R_G+R_E).
\label{eq:finitebound}
\end{equation}
Thus, it is the number of independent Schmidt branches generated by the
participating configurations, rather than their number alone, that controls
the entanglement across a given bipartition.

The optimization crossings considered previously lie in the dilute-manifold
regime, with $\eta_G,\eta_E\to0$ exponentially and only $O(1)$
configurations in the relevant low-energy sectors as $L$ grows. Hence
$R_G,R_E=O(1)$ and the Schmidt structure remains bounded even when the
competing manifolds are separated by an extensive Hamming distance. Writing
the relevant inter-manifold distance as $d_H\simeq cL$, with $0<c\leq1$,
gives
\begin{equation}
\Delta_{\min}\sim(1-s)^{cL},
\end{equation}
up to energy-denominator and path-counting factors. Here $c$ measures the
fraction of spins that must reorganize between the competing manifolds, with
$c=1$ corresponding to maximal Hamming separation. This is the regime
realized by the structured and random optimization examples.

To obtain a growing Schmidt rank, the dilute-manifold picture must therefore
break down. Configurations differing by a single spin cannot generate an
independent Schmidt branch across a bipartition, since the changed spin
belongs entirely to one subsystem (cf.\ Sec.~\ref{sec:twostate}, where the
same coefficient-matrix argument shows that a single admixed configuration
only redistributes existing Schmidt weight). The smallest distance capable
of generating a new branch is therefore $d_H=2$, with the two flipped spins
lying on opposite sides of the cut. Such configurations first hybridize at
second order, 
$t\sim(1-s)^2$, for finite energy denominators. Near the end of the anneal, where
$s\rightarrow1$, appreciable hybridization consequently requires
near-degeneracies on the scale of this matrix element. A growing Schmidt
structure thus requires an increasing number of low-energy configurations
that can hybridize within the same manifold, rather than isolated
perturbative resonances.

Ground-state degeneracy provides one natural route to such a non-dilute
manifold. Many constraint and optimization problems, including MaxSAT,
XORSAT, and graph colouring, can possess exponentially large sets of
degenerate optimal solutions~\cite{BAPST2013127}. If an $O(L)$ or larger
number of these configurations contribute to the annealing ground state,
they can generate a growing Schmidt structure. However, in this case the
same manifold remains the ground sector as $s\to1$, so the resulting
entanglement is directly tied to the degeneracy of the problem ground
space and does not by itself separate the origin of the Schmidt structure
from the endpoint bottleneck.

More generally, the $O(L)$ configurations responsible for the growing
Schmidt structure need not belong to the ground-state manifold of the
problem Hamiltonian. A competing excited sector can itself contain an
increasing number of low-energy configurations that hybridize through
low-order processes. Although the problem ground state may then be
nondegenerate, the dressed ground state at an avoided crossing can acquire
a contribution from this internally hybridized excited sector and thereby
develop a growing Schmidt rank. This suggests that the relevant
construction should allow both competing low-energy sectors to develop
nontrivial superpositions, while keeping their mutual coupling
exponentially small.

The simplest realization is therefore to construct two competing
low-energy manifolds, each containing an increasing number of
configurations that are connected by low-order processes, while the
distance between the two manifolds remains extensive in Hamming space.
The resulting separation between intra-manifold hybridization and
inter-manifold tunnelling allows the Schmidt structure and the spectral
bottleneck to arise from parametrically different processes.

It is useful first to ask how many mutually separated configurations such a
structure can contain. Let $R(L)$ denote the number of computational
configurations in such a manifold, viewed as a binary code of length $L$ and
minimum Hamming distance $d$. Requiring mutual separation $d=L/2$ places
these configurations exactly at the equality case $2d=L$ of the Plotkin
bound. At this boundary the general ratio bound
$M\le 2\lfloor d/(2d-L)\rfloor$ diverges, but the equality case itself is
bounded separately: $M\le4d$~\cite{1057584}. Hence
\begin{equation}
R(L)\leq4d=2L,
\qquad
R(L)=O(L),
\label{eq:plotkin}
\end{equation}
and correspondingly the manifold density is at most
\begin{equation}
\eta(L)\lesssim\frac{L}{2^L},
\end{equation}
which remains exponentially small even in the most favorable geometric
arrangement. Consequently, a manifold whose configurations are all
separated by $L/2$ can support 
$S\lesssim\log R(L)\sim\log L$ 
bipartite entropy. In particular, an exponentially small gap generated by
tunnelling across an extensive Hamming distance does not allow an
arbitrarily large number of mutually separated configurations to contribute
to the same low-energy manifold. This is a geometric constraint on a set of
mutually separated configurations, rather than a general entanglement
bound: configurations within a low-energy manifold need not remain mutually
distant, and their Schmidt rank can be smaller than their number.

For intuition, at $L=8$ one can explicitly construct eight configurations,
\begin{equation}
\begin{aligned}
&\ket{0000|0000},\quad
\ket{0010|1110},\quad
\ket{0101|0110},\quad
\ket{0111|1000},\\
&\ket{1001|1010},\quad
\ket{1011|0100},\quad
\ket{1100|1100},\quad
\ket{1110|0010}.
\end{aligned}
\label{eq:L8code}
\end{equation}
Every pair is separated by $L/2=4$ spins, while all left and right
substrings are distinct. Their equal-weight superposition therefore has
Schmidt rank $r_S=8$ and entropy
\begin{equation}
S=\ln8=\ln L
\end{equation}
across the $4|4$ cut. This provides a simple kinematic illustration that
logarithmically growing Schmidt rank is compatible with extensive mutual
Hamming separation. However, realizing this structure dynamically would
require several configurations to become simultaneously relevant at low
energy, rather than the two-sector avoided crossing considered in standard
annealing bottlenecks. Moreover, these configurations do not form a
low-order perturbative manifold: connecting them through a transverse-field
driver requires extensive-order processes. Thus, while the construction
demonstrates the geometric possibility, engineering a Hamiltonian that
naturally realizes such a multi-state resonance is nontrivial.

We instead seek a more controlled two-manifold construction. The two
competing low-energy manifolds should each contain an increasing number of
configurations that are connected by low-order processes, so that they
develop a nontrivial internal superposition and growing Schmidt structure.
At the same time, every configuration in one manifold should remain
extensively separated in Hamming space from those in the other, keeping
the inter-manifold tunnelling exponentially suppressed. In this way, the
process that generates the Schmidt structure is parametrically distinct
from the process that produces the spectral bottleneck. This leads to the
\emph{mirror-twin manifold} construction.

Consider $L=2n$ data spins divided into equal subsystems $A$ and $B$, together
with one selector spin $C$. Define
\begin{equation}
x_k=
(\underbrace{\uparrow\cdots\uparrow}_{k}
\underbrace{\downarrow\cdots\downarrow}_{n-k}),
\qquad k=0,\ldots,n-1,
\end{equation}
and the two manifolds
\begin{equation}
\mathcal G=
\left\{
\ket{x_k}_A\ket{x_k}_B\ket{\uparrow}_C
\right\}_{k=0}^{n-1},
\qquad
\mathcal E=
\left\{
\ket{\bar{x}_k}_A\ket{x_k}_B\ket{\downarrow}_C
\right\}_{k=0}^{n-1},
\end{equation}
where $\bar{x}_k$ is obtained by flipping every spin of $x_k$. Neighbouring
configurations within either manifold differ by one spin in $A$ and one in
$B$, so that
$d_H(g_k,g_{k+1})
=
d_H(e_k,e_{k+1})
=
2$.
In contrast, any pair drawn from different manifolds satisfies $d_H(g_k,e_j)=n=\frac{L}{2}$,
when the selector spin is excluded from the Hamming distance. Each manifold
contains $n=L/2$ configurations, so $R_G=R_E=L/2$ and $\eta_G=\eta_E=\frac{L}{2^{L+1}}$, which decays exponentially with $L$ but contains a number of configurations
that grows linearly with system size.

The two manifolds are realized as low-energy sectors of the diagonal,
three-local problem Hamiltonian
\begin{equation}
H_P=
(U_0+U_1Z_C)\,P_B
+\frac K2\sum_{i=1}^{n}
\left(1-Z_i^A Z_i^B Z_C\right)
+\frac{\delta}{2}(1-Z_C),
\label{eq:designerHP}
\end{equation}
where
\begin{align}
P_B&=
\frac{1+Z_n^B}{2}
+\frac14\sum_{i=1}^{n-1}
(1-Z_i^B)(1+Z_{i+1}^B), \nonumber \\
U_{0,1}&=\frac{U_G\pm U_E}{2}.
\end{align}
The penalty $P_B$ vanishes precisely on the $n$ configurations $x_k$. The
three-spin terms enforce $Z_i^A=Z_i^B$ for $Z_C=+1$ and
$Z_i^A=-Z_i^B$ for $Z_C=-1$, thereby selecting $\mathcal G$ and
$\mathcal E$, respectively. Thus the two manifolds have classical energies
$0$ and $\delta$, while $U_G$ and $U_E$ control their virtual excursions away
from the allowed $B$ configurations.

We anneal this problem with the standard transverse-field driver,
\begin{equation}
H(s)=
-(1-s)\Gamma
\left(
\sum_{i=1}^{n}X_i^A+
\sum_{i=1}^{n}X_i^B+
X_C
\right)
+sH_P .
\label{eq:designer_anneal}
\end{equation}
Within either manifold, neighbouring configurations are connected at second
order through intermediate states that violate the three-spin constraint.
Thus the effective intra-manifold hopping scales as
\begin{equation}
t_\alpha\sim
\frac{(1-s)^2\Gamma^2}{sK_\alpha},
\qquad \alpha=G,E,
\end{equation}
up to numerical factors, where $K_\alpha$ denotes the corresponding
excitation energy. Each manifold consequently develops a delocalized
low-energy state over $O(n)$ configurations. Across the
$(A,C)|B$ bipartition, the $B$ strings $x_k$ are distinct, so the resulting
Schmidt rank scales as
\begin{equation}
r_S=\Theta(n)=\Theta(L),
\qquad
S_{AC|B}=\Theta(\log L).
\end{equation}
This is by design: the `kink' is the domain wall between the
$\uparrow$ and $\downarrow$ regions of $x_k$. The same domain-wall
position $k$ occurs in both $A$ and $B$, so each value of $k$ gives a
distinct Schmidt branch across $AC|B$. Across $A|BC$ the rank is also $2n$
(Appendix~\ref{app:appC}), so $r_S=\Theta(L)$ for either cut.

The offset $\delta$, together with the different virtual corrections
controlled by $U_G$ and $U_E$, can be tuned to bring the two dressed
manifolds into resonance. At the resulting avoided crossing, the ground
state is a coherent superposition of the two internally hybridized
manifold states. Yet a configuration in $\mathcal G$ must still undergo
$n=L/2$ data-spin flips to reach any configuration in $\mathcal E$.
Including the selector requires one additional flip, so the inter-manifold
matrix element occurs at order $L/2+1$ in the transverse-field amplitude,
\begin{equation}
t_{\rm inter}
\sim
[(1-s)\Gamma]^{L/2+1},
\end{equation}
up to energy-denominator and path-counting factors. Equivalently, after
factoring out the relevant energy scale, its dimensionless perturbative
parameter is of order
\[
\left(\frac{(1-s)\Gamma}{sK}\right)^{L/2+1},
\]
up to corrections from the different intermediate-state energies. The
additional selector flip changes only the prefactor in the exponential
scaling with $L$.

Exact diagonalization confirms the resulting separation of scales. As shown
in Fig.~\ref{fig:designed}, the minimum gap decreases exponentially with
system size, while the entropy across $(A,C)|B$ grows logarithmically,
\begin{equation}
\Delta_{\min}\sim e^{-1.017L},
\qquad
S_{AC|B}\sim\ln L.
\end{equation}
Thus the construction realizes the desired regime in which the low-energy
state is delocalized over $O(L)$ configurations within each competing
manifold, while the coupling between the manifolds remains exponentially
suppressed.
The degrees of freedom that generate the Schmidt structure are also those
that mediate the bottleneck: every data spin is flipped in passing from
$\mathcal G$ to $\mathcal E$, while the same spins encode the kink position
and generate the Schmidt branches across the cut. The selector $C$ is
likewise not a spectator, since it contributes an additional flip to the
inter-manifold tunnelling process. The separation between gap and
entanglement therefore does not arise from assigning these properties to
different subsystems, but from the hierarchy of perturbative processes:
low-order processes hybridize configurations within each manifold, whereas
extensive-order processes connect the two manifolds. The construction is
thus not a product of a hard sector with a decoupled entangled one. Additionally, the construction is non-dilute in the sense that the number of
configurations in each low-energy manifold grows linearly with system size.
The resulting separation of perturbative scales provides an explicit
realization of growing Schmidt structure together with an exponentially
suppressed inter-manifold bottleneck.
\subsection{Overlap-driven tunneling and the antiferromagnetic Ising chain}
\label{sec:overlaptunnel}
\begin{figure}
    \centering
    \subfloat[]{%
        \includegraphics[width=0.45\linewidth]{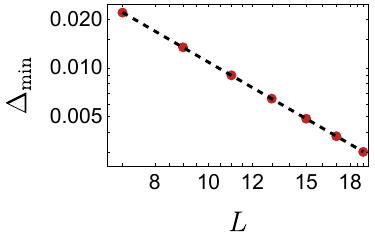}
    }\hfill
    \subfloat[]{%
        \includegraphics[width=0.45\linewidth]{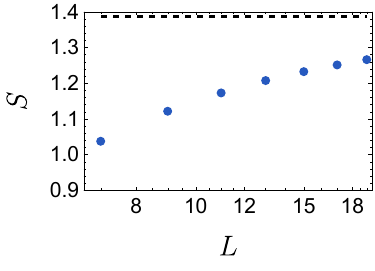}
    }
    \caption{
    Finite-size scaling at the first-order transition of the odd-length
    antiferromagnetic Ising chain. 
    (a) The minimum spectral gap $\Delta_{\min}$ as a function of system
    size $L$ on a log-log scale. The dashed line shows the power-law fit
    $\Delta_{\min}\sim L^{-1.98}$. 
    (b) The bipartite entanglement entropy $S$ of the ground state,
    approaching the finite asymptotic value $\log 4$ (dashed line) as
    $L$ increases. Despite the polynomially closing gap, the
    Schmidt rank remains bounded by four.
    }
    \label{fig:AFMising}
\end{figure}
The perturbative picture above is not the only mechanism by which an
exponentially small gap can arise. The $p$-spin model considered in
Sec.~\ref{sec:sec2} provides a qualitatively different example. At its
first-order transition, the two competing low-energy states evolve from the
ground states of the transverse-field driver and the problem Hamiltonian,
respectively. The ground state of the transverse-field driver is not an
individual computational-basis configuration, but a coherent many-body
state with support on a large number of configurations. Nevertheless, the
matrix element coupling the two competing states becomes exponentially small
with system size, resulting in an exponentially closing gap. Thus,
exponential suppression of the gap can arise not only from high-order
tunnelling processes between distant computational-basis configurations,
but also from weak coupling between distinct many-body states. This
distinction will be important when we consider low-energy manifolds
containing many configurations.

For the fully connected $p$-spin model, the first-order transition for odd
$p>2$ occurs between a paramagnetic branch and a ferromagnetic branch. Near
the transition, for odd $p$ these can be approximated by
$\ket{P}=\ket{+}^{\otimes L}
=2^{-L/2}\sum_z\ket z$ and
$\ket{F}\simeq\ket{\uparrow}^{\otimes L}$, where $\ket{z}$ denotes a computational basis state.
The effective coupling between the two branches is
$t_{\rm eff}=\bra{P}H(s_c)\ket{F}$, with
\begin{equation}
t_{\rm eff}
=
-(1-s_c)
\bra{P}\sum_i\sigma_i^x\ket{F}
+
s_c\bra{P}H_P\ket{F}.
\label{eq:teffpspin}
\end{equation}
Since $H_P\ket{F}=E_F\ket{F}$ with $E_F=O(L)$,
$\bra{P}H_P\ket{F}=E_F\braket{P|F}\sim L\,2^{-L/2}$, while
$\bra{P}\sum_i\sigma_i^x\ket{F}=L\,2^{-L/2}$. Thus, up to model-dependent
prefactors, $t_{\rm eff}\sim L\,2^{-L/2}$ and
$\Delta_{\min}\sim L\,2^{-L/2}$.

The exponential suppression therefore has a different origin from the
perturbative tunneling discussed above: it arises from the exponentially
small overlap between a delocalized paramagnetic phase and a localized
ordered branch, rather than from an extensive perturbative order in the
transverse-field driver.
The entanglement structure of the $p$-spin transition remains bounded
alongside this exponentially small gap. For odd $p>2$, the critical state is
approximately
\[
\ket{\psi}\simeq a\ket{P}+b\ket{F},
\]
giving $S\simeq\log2$ across an equal bipartition, since both $\ket{P}$ and
$\ket{F}$ essentially have Schmidt rank $r_S=1$. For even $p$, the
ferromagnetic phase is symmetry-degenerate,
\[
\ket{F}\simeq
\frac{\ket{\uparrow}^{\otimes L}+\ket{\downarrow}^{\otimes L}}{\sqrt{2}},
\]
so that the critical state contains three independent branches and
$S\simeq\log3$. More generally, if the ordered sector contains $R$ distinct
collective branches
$\mathcal M_F=\{\ket{F_\alpha}\}_{\alpha=1}^{R}$, the critical state can be
written as
\[
\ket{\psi}
=
a\ket{P}
+
\sum_{\alpha=1}^{R}b_\alpha\ket{F_\alpha},
\]
giving
\begin{equation}
r_S = 1+R,
\qquad
S\leq\log(1+R).
\label{eq:pspinSchmidtbound}
\end{equation}
Thus, increasing the degeneracy of the ordered sector can increase the
Schmidt rank, but only through the number of independent branches that
participate in the critical state. This is the same mechanism encountered
for exponentially degenerate solution manifolds such as those discussed for
highly-degenerate MaxSAT problems in the previous section, although the
mechanism responsible for the exponentially closing gap during annealing
is different. In both cases, the scaling of the entanglement is controlled
by the independent branches of the degenerate sector rather than directly
by the mechanism producing the small gap. The $p$-spin model therefore
illustrates how an exponentially small gap can coexist with bounded
entanglement even when the small matrix element is not generated by a
high-order sequence of local spin flips.

This naturally raises the opposite question: can a first-order transition
possess an extensive low-energy manifold while exhibiting only a polynomially
closing gap and bounded Schmidt rank? An explicit realization is provided
by the \emph{staggered-field antiferromagnetic Ising chain} studied in
Ref.~\cite{PhysRevLett.109.030502}:
\begin{equation}
H=J\sum_i\sigma_i^z\sigma_{i+1}^z
-\sum_i(-1)^i h\,\sigma_i^z
-\Gamma\sum_i\sigma_i^x.
\end{equation}

For $\Gamma<J$, sweeping the staggered field through $h=0$ produces a
first-order transition between the two N\'eel-ordered phases. Consider an
odd-length periodic chain. At $h=0$, perfect N\'eel order is frustrated, so
every lowest-energy configuration contains a single frustrated bond. The
frustrated bond can occupy any of the $L$ bonds, while the two possible
orientations of the surrounding N\'eel pattern give a $2L$-fold low-energy
manifold.

This extensive manifold has a qualitatively different connectivity from the
perturbative avoided crossings considered above. A single application of the
transverse-field term moves the frustrated bond by one lattice spacing.
Within the low-energy manifold, the driver therefore acts at first order as
nearest-neighbor hopping on a ring of length $2L$. The effective Hamiltonian
can be written as
\begin{equation}
H_{\rm eff}
=
E_0
-\Gamma_{\rm eff}
\sum_{j=1}^{2L}
\left(
\ket{j+1}\bra j+\ket j\bra{j+1}
\right),
\label{eq:AFMeff}
\end{equation}
with periodic boundary conditions and $\Gamma_{\rm eff}=O(\Gamma)$. The
resulting levels are those of a particle on a ring,
\begin{equation}
E_q=E_0-2\Gamma_{\rm eff}\cos\left(\frac{\pi q}{L}\right),
\end{equation}
so that
\begin{equation}
\Delta_{\min}
=
2\Gamma_{\rm eff}
\left[
1-\cos\left(\frac{\pi}{L}\right)
\right]
\simeq
\frac{\pi^2\Gamma_{\rm eff}}{L^2}.
\label{eq:AFMgap}
\end{equation}
Thus, despite the first-order nature of the transition, the gap is
polynomial rather than exponentially small. This scaling is confirmed by
the exact diagonalization shown in Fig.~\ref{fig:AFMising}(a), where the
numerical data give
$\Delta_{\min}\sim L^{-1.98}$, consistent with the expected $L^{-2}$
behavior.

The extensive number of configurations in this manifold does not,
however, imply an extensive Schmidt rank. To see this explicitly, consider
a bipartition into left and right halves. The crucial point is that the
different positions of the frustrated bond do not generate independent
Schmidt branches. If the frustrated bond lies in the left half, all such
configurations have the \emph{same} N\'eel configuration on the right,
apart from the choice of the two global N\'eel orientations. Conversely, if
the frustrated bond lies in the right half, all such configurations have
the same N\'eel configuration on the left, again for each of the two
orientations.

For a bipartition of $\lfloor L/2\rfloor$ and $\lceil L/2\rceil$, let $A$ and $B$ denote the left and right
subsystems, respectively. The equal-weight ground state within the
low-energy manifold can then be grouped schematically as
\begin{equation}
\ket{\psi_0}
=
\sum_{\alpha=\pm}
\left[
\sum_{j\in A}
\ket{\ell_{\alpha,j}}_A
\ket{R_\alpha}_B
+
\ket{L_\alpha}_A
\sum_{j\in B}
\ket{r_{\alpha,j}}_B
\right].
\end{equation}
Here $\alpha=\pm$ labels the two possible global N\'eel orientations,
while $j$ labels the position of the frustrated bond. For $j\in A$,
$\ket{\ell_{\alpha,j}}_A$ denotes the left-subsystem configuration with
the frustrated bond at position $j$, while $\ket{R_\alpha}_B$ is the
corresponding N\'eel configuration of the right subsystem. Conversely,
for $j\in B$, $\ket{r_{\alpha,j}}_B$ denotes the right-subsystem
configuration with the frustrated bond at $j$, while
$\ket{L_\alpha}_A$ is the corresponding N\'eel configuration of the left
subsystem. The precise normalization is not important for the Schmidt-rank
counting. The sums over all defect positions on one
side of the cut therefore combine into a \emph{single} Schmidt vector on
that side, because the state of the opposite subsystem is identical for
all those positions. There are consequently at most two branches
associated with the defect being on the left, and two associated with it
being on the right:
\begin{equation}
r_S= 2\times2=4,
\qquad
S\leq\ln4.
\label{eq:AFMentropy}
\end{equation}
Thus the $2L$ configurations in the low-energy manifold do not produce
$2L$ independent Schmidt coefficients. The extensive manifold describes
many configurations within the same few Schmidt sectors. In the large-$L$
limit all four sectors are populated, and the bound is saturated,
$S\rightarrow\ln4$, as confirmed by the numerical results in
Fig.~\ref{fig:AFMising}(b).

The staggered-field antiferromagnetic Ising chain therefore provides a
particularly transparent counterexample to any direct identification of
first-order transitions with exponentially small gaps or growing
entanglement. It has a first-order transition and an extensive
$2L$-dimensional low-energy manifold, yet its gap closes polynomially and
its Schmidt rank remains bounded. More generally, such behavior can arise
when a low-energy manifold is generated by a bounded number of localized
defects, whose origin may be frustration, boundary conditions, as well as
topological or other local constraints. The bounded defect content limits
the number of independent Schmidt sectors, while the driver must also
connect the defect configurations at low order so that the manifold can
hybridize coherently. In the present case, frustration produces a single
mobile `kink', a frustrated bond, and the transverse field moves it at
first order, yielding the $L^{-2}$ level spacing of the resulting
dispersive band. This should be contrasted with the mirror-twin manifold
of Sec.~\ref{sec:mirrortwin}, where the analogous kink is tied to a second
kink in the opposite subsystem: each kink position is then reproduced
independently on both sides of the cut and generates its own Schmidt
branch, whereas here the subsystem opposite the kink is always the same
uniform N\'eel background. Hence,  only which side carries the kink and not its
exact position, is visible across the cut. It is this difference in how
the defect label is encoded across the bipartition, rather than merely
the hopping order, that separates the bounded entanglement found here
from the growing entanglement of the mirror-twin construction. Thus, a
large low-energy manifold need not imply either large entanglement or an
exponentially small gap: what matters is both the internal complexity of
the configurations and their connectivity under the driver. This
bounded-defect picture is precisely what breaks down at a continuous
transition, where low-energy fluctuations proliferate across increasing
length scales.
\subsection{Continuous transitions and growing low-energy manifolds}

The first-order transitions considered above illustrate how an exponentially
small gap can arise from the coexistence and weak hybridization of distinct
low-energy sectors. The detailed structure of these sectors can vary
substantially, as demonstrated by the counterexamples constructed above:
neither the gap scaling nor the Schmidt rank is fixed solely by the presence
of a first-order transition. A qualitatively different situation arises at a
continuous transition, where the distinction between the competing phases
vanishes at the critical point and the ground state evolves continuously
between them.

A typical feature of a continuous transition is a polynomially closing gap.
Compared with an exponentially closing gap, this represents a parametrically
weaker suppression of the gap with system size. A concrete example is provided by the $p=2$ member of the fully connected
$p$-spin model considered in Sec.~\ref{sec:sec2}, equivalently the
Lipkin--Meshkov--Glick (LMG) model. In contrast to the first-order
transitions of the $p>2$ models, the $p=2$ model undergoes a continuous
transition. As shown in Fig.~\ref{fig:pspin}, the minimum gap closes
polynomially with system size, while the bipartite entanglement grows
logarithmically,
\begin{equation}
\Delta_{\min}\sim L^{-1/3},
\qquad
S\sim\log L.
\end{equation}
Here there is no analogue of the bounded single-defect description of the
staggered-field antiferromagnet. Instead, the critical state is dressed by an
increasing number of collective fluctuations, producing a growing Schmidt
structure.

More generally, the bounded-defect picture discussed above breaks down at a
continuous transition: low-energy fluctuations become relevant on
increasingly large length scales as the critical point is approached. Rather
than being described by a finite number of defect sectors, the ground state
is continuously dressed by fluctuations spanning a growing range of length
scales. In spatially extended systems, these fluctuations can be viewed in
terms of domains and domain-wall configurations of the competing order,
while in fully connected models such as the $p=2$ LMG model they instead
take the form of collective fluctuations. The common feature is that the
low-energy state can no longer be organized in terms of a bounded number of
defect configurations or independent sectors.

This change is reflected directly in the low-energy spectrum. Near a
continuous transition, the ground state is not isolated from a fixed number
of competing branches. Instead, a growing set of low-lying excitations
becomes relevant as the system size increases, with their characteristic
energy scales governed by critical finite-size scaling. The critical ground
state consequently involves an increasing number of correlated fluctuations
and cannot be organized into a finite set of independently hybridizing
low-energy sectors. This is qualitatively different from the finite-branch
avoided crossings considered above, even though both can produce bottlenecks
in the annealing spectrum.

The growing low-energy structure is also reflected in bipartite
entanglement. For one-dimensional critical systems described by a conformal
field theory, the entanglement entropy of an interval grows logarithmically
with its size~\cite{Calabrese2009},
\begin{equation}
S=\frac{c}{3}\log L+\mathrm{const.},
\end{equation}
for periodic boundary conditions, with the corresponding coefficient
modified by the boundary conditions and bipartition geometry. Thus, in the
models considered here, the polynomially closing gap is accompanied by a
Schmidt rank that grows with system size rather than remaining confined to a
finite number of branches. Continuous quantum criticality can also support
genuine multipartite correlations, providing a complementary manifestation
of the same growing many-body structure~\cite{PhysRevB.101.054431}.

It is important, however, that the growth of the wavefunction's support in
the computational basis is not itself the origin of the entanglement. The
paramagnetic state
\begin{equation}
\ket{+}^{\otimes L}
=
2^{-L/2}\sum_{\boldsymbol{\sigma}}
\ket{\boldsymbol{\sigma}}
\end{equation}
already has support on all $2^L$ computational-basis configurations while
having Schmidt rank one across every bipartition. What matters instead is
the structure of the correlations between the two subsystems. 
At a
continuous transition, fluctuations on increasingly large length scales
generate an increasing number of independent Schmidt components.

The contrast with the first-order cases can therefore be stated in terms of
both connectivity and internal structure of the low-energy manifold. In the
first-order bottlenecks considered above, an exponentially small gap can
result when distinct low-energy sectors are coupled only through
high-order processes, while the configurations within a sector may retain
bounded or growing Schmidt structure independently of this inter-sector
tunnelling. In the staggered-field antiferromagnet, by contrast, a bounded
defect structure is strongly hybridized at low order, producing a
polynomial gap without growing entanglement. The mirror-twin manifold of Sec.~\ref{sec:mirrortwin} occupies the remaining combination: a non-dilute resonance hybridizes a growing number of kink configurations at low order within each manifold, while the two manifolds themselves remain separated by an extensive Hamming distance, so a growing Schmidt structure coexists with an exponentially suppressed gap. At a continuous transition,
this bounded-structure picture breaks down: the relevant fluctuations
proliferate across increasing length scales, producing a growing set of
independent Schmidt components. The polynomial gap closing and logarithmic
entanglement therefore arise together from the critical low-energy
structure, rather than from a single tunnelling matrix element between a
small number of competing branches.
\section{Discussion}
\label{sec:sec4}

Our results distinguish the complexity of an annealing state from the
difficulty of accessing it. Exponentially small gaps can coexist with
$O(1)$ bipartite entanglement, while extensive low-order hybridization can
modify the gap without generating a correspondingly complex Schmidt
structure; continuous transitions, by contrast, exhibit polynomially
closing gaps together with logarithmically growing entanglement. The gap
and entanglement therefore probe distinct aspects of the low-energy
landscape.

As shown above, this distinction traces back to the structure and
connectivity of the low-energy manifold, rather than to the order of the
transition itself. The mirror-twin construction of
Sec.~\ref{sec:mirrortwin} completes this picture: by hybridizing a growing
number of configurations at low order within each of two manifolds held at
extensive Hamming separation, it realizes the remaining combination,
exponential gap with growing entanglement, not obtained in the examples of
Sec.~\ref{sec:sec2} or \ref{sec:sec3}. The gap and entanglement can
therefore scale independently, realizing all four combinations summarized
in Table~\ref{tab:gap_entanglement}.

This distinction is also relevant to the role often attributed to
multipartite entanglement in quantum annealing. Multipartite entanglement
has been proposed both as a resource and as a diagnostic of quantum
annealing performance~\cite{Batle2016,Santra2025}. However, multipartite
correlations are not unique to exponentially difficult first-order
transitions: continuous quantum phase transitions also support genuine
multipartite entanglement~\cite{Hauke2016,PhysRevB.101.054431}, including
the $p=2$ transition considered here. Thus, the presence of multipartite
entanglement alone does not identify the quantum resource responsible for
efficient annealing.

This raises a more fundamental question: if the spectral gap characterizes
the difficulty of an annealing path, but is itself not a quantum resource,
what quantum resource is actually being exploited when annealing succeeds?
Likewise, if entanglement complexity does not determine the gap scaling,
what property of the evolving quantum state or of its trajectory through
Hilbert space must be controlled to achieve efficient state preparation?
Our results suggest that the answer may be intrinsically geometric. Rather
than asking only how entangled the instantaneous state is, one may ask how
the state traverses Hilbert space as the annealing parameter is varied, and
what geometric structure governs its accessibility.

The quantum geometric tensor provides a natural starting point for such a
description~\cite{ProvostVallee1980,Gu2010}. Its metric component measures
the distinguishability of nearby ground states as the annealing parameter
is varied, and therefore characterizes the geometry of the trajectory
through Hilbert space. This perspective also connects naturally to the
quantum Fisher information, but here an important distinction is needed.
For a pure ground state parameterized by the annealing parameter $s$, the
corresponding parameter-space QFI is $F_Q^{(s)}=4g_{ss}$.
This quantity measures the sensitivity of the state to motion along the
annealing path. It is distinct from the QFI with respect to a physical
generator, which has been widely used as a measure or witness of
multipartite entanglement~\cite{Hauke2016}. The two quantities therefore
probe different aspects of the state: the former characterizes the geometry
of the annealing trajectory, while the latter characterizes the response to
a chosen physical observable. This distinction is particularly relevant
here, since our results show that entanglement complexity alone does not
capture the structure responsible for the annealing bottleneck.

Whether a geometric quantity derived from the quantum geometric tensor, or
some more general geometry-based quantum resource, can identify what is
effectively optimized during efficient annealing remains an open question.
In particular, the quantum metric is strongly enhanced near small gaps, so
it is not yet clear whether it contains information beyond the spectral gap
itself. A useful geometric resource would instead need to capture the
Hilbert-space accessibility and connectivity revealed here, while providing
information not already contained in either the gap or entanglement alone.
\begin{acknowledgments}
R.G. and S.B. thank P. A. Warburton for discussions. We acknowledge the supports of UK EPSRC project EP/Y004590/1 and EPSRC-SFI project EP/X039889/1 during the initial stages of this work.
\end{acknowledgments}

\bibliography{references}
\appendix

\section{Bipartite entanglement from the permutation-symmetric sector}
\label{app:entanglement}

The fully connected $p$-spin Hamiltonian is permutation symmetric and therefore
preserves the maximum-spin sector
$\mathcal{S}=L/2$. This allows the spectral problem to be reduced from the
full Hilbert space of dimension $2^L$ to the Dicke subspace of dimension
$L+1$. The bipartite entanglement entropy, however, cannot be obtained by
simply treating this reduced space as a tensor product of the two physical
subsystems. We therefore reconstruct the bipartite coefficient matrix using
the Clebsch--Gordan decomposition of the collective spin.

We consider an equal bipartition of the $L$ spins into subsystems $A$ and
$B$, containing $L_A$ and $L_B=L-L_A$ spins, respectively. The corresponding
maximum-spin sectors are
\[
\mathcal{S}_A=\frac{L_A}{2},
\qquad
\mathcal{S}_B=\frac{L_B}{2}.
\]
A state in the fully symmetric sector can be written in the Dicke basis as
\begin{equation}
|\psi\rangle
=
\sum_{m=-L/2}^{L/2}
c_m\,\bigl|\mathcal{S}=L/2,m\bigr\rangle .
\label{eq:app_dicke_state}
\end{equation}
Here the coefficients $c_m$ are obtained by diagonalizing the reduced
$(L+1)$-dimensional Hamiltonian.

Although the spectral problem can be solved entirely within the
permutation-symmetric sector, the bipartition used to define the
entanglement entropy is a bipartition of the \emph{physical spins}.
Specifically, we divide the $L$ spins into two sets, $A$ and $B$, with
$L_A$ and $L_B$ spins, respectively. Thus, the physical Hilbert space
factorizes as
\[
\mathcal{H}=\mathcal{H}_A\otimes\mathcal{H}_B,
\]
and the reduced density matrix and entanglement entropy are defined with
respect to this tensor-product structure.

The permutation-symmetric sector is not itself a tensor product
$\mathcal{H}_{\mathcal S}\neq
\mathcal{H}_{\mathcal S_A}\otimes\mathcal{H}_{\mathcal S_B}$.
Instead, when a fully symmetric state of all $L$ spins is viewed across
the physical bipartition $A|B$, it can be decomposed into symmetric
states of the two subsystems. The corresponding collective spins are
\[
\mathcal{S}_A=\frac{L_A}{2},\qquad
\mathcal{S}_B=\frac{L_B}{2},
\]
and their addition gives the maximum total spin
\[
\mathcal{S}=\mathcal{S}_A+\mathcal{S}_B=\frac{L}{2}.
\]
Consequently, the Dicke states of the full system can be expanded in the
product basis of the subsystem Dicke states using Clebsch-Gordan
coefficients.

\begin{equation}
\bigl|\mathcal{S},m\bigr\rangle
=
\sum_{m_A,m_B}
\left\langle
\mathcal{S}_A,m_A;\mathcal{S}_B,m_B
\middle|
\mathcal{S},m
\right\rangle
\bigl|\mathcal{S}_A,m_A\bigr\rangle
\otimes
\bigl|\mathcal{S}_B,m_B\bigr\rangle ,
\label{eq:app_CG}
\end{equation}
where the Clebsch--Gordan coefficient enforces
\[
m=m_A+m_B.
\]
Thus, the Clebsch-Gordan transformation does not define a new
bipartition; rather, it expresses the permutation-symmetric basis of the
full system in the tensor-product basis associated with the physical
bipartition $A|B$. This restores precisely the tensor-product structure
needed to construct the reduced density matrix.
Substituting Eq.~\eqref{eq:app_CG} into Eq.~\eqref{eq:app_dicke_state}
gives
\begin{equation}
|\psi\rangle
=
\sum_{m_A,m_B}
M_{m_A,m_B}
\bigl|\mathcal{S}_A,m_A\bigr\rangle
\otimes
\bigl|\mathcal{S}_B,m_B\bigr\rangle ,
\label{eq:app_coeffmatrix}
\end{equation}
with coefficient matrix
\begin{equation}
M_{m_A,m_B}
=
c_{m_A+m_B}
\left\langle
\mathcal{S}_A,m_A;\mathcal{S}_B,m_B
\middle|
\mathcal{S},m_A+m_B
\right\rangle .
\label{eq:app_M}
\end{equation}
Thus, although the ground state is obtained in the reduced collective-spin
basis, its bipartite Schmidt decomposition is recovered from the singular
value decomposition of the matrix $M$.

Writing
\begin{equation}
M=U\Lambda V^\dagger,
\end{equation}
the nonzero singular values $\lambda_\alpha$ are the Schmidt coefficients.
The reduced density matrix eigenvalues are consequently
$p_\alpha=\lambda_\alpha^2$, and the bipartite von Neumann entropy is
\begin{equation}
S
=
-\sum_{\alpha} \lambda_\alpha^2
\ln\lambda_\alpha^2 .
\label{eq:app_entropy}
\end{equation}
The coefficients are normalized such that
$\sum_\alpha\lambda_\alpha^2=1$.

For each system size, we diagonalize the Hamiltonian within the maximum-spin
sector and determine the minimum instantaneous gap along the annealing path.
The ground-state eigenvector at the minimum-gap point is then used in
Eq.~\eqref{eq:app_M} to construct the bipartite coefficient matrix. The
entropy is evaluated from its singular values. We use an equal bipartition,
$L_A=L_B=L/2$, throughout the main text.

For even $p$, the Hamiltonian also possesses the global spin-flip symmetry
$m\rightarrow-m$. In the numerical calculation we work in the corresponding
even-parity sector when evaluating the ground state. Explicitly, the Dicke
basis states with $m$ and $-m$ are combined as
\begin{equation}
|m\rangle_{+}
=
\frac{1}{\sqrt{2}}
\left(
|\mathcal{S},m\rangle+
|\mathcal{S},-m\rangle
\right),
\qquad m>0,
\label{eq:app_parity}
\end{equation}
with the $m=0$ state included separately when present. The resulting
Hamiltonian is diagonalized in this symmetry-reduced basis, after which the
eigenvector is transformed back to the Dicke basis before constructing
$M$. This last step is essential because the Clebsch--Gordan decomposition
in Eq.~\eqref{eq:app_CG} is defined in the Dicke basis.

The procedure therefore separates the two reductions used in the calculation:
permutation symmetry reduces the spectral problem to a space of dimension
$L+1$, while the Clebsch--Gordan decomposition restores the tensor-product
structure required to calculate the physical bipartite entanglement.
\section{Bipartite graph}
\label{app:appA}

We briefly summarize the bipartite-graph construction used in this work;
the detailed derivation and its connection to the maximum-weight independent
set problem are given in Ref.~\cite{hnzj-j7kt}.

We consider a complete bipartite graph with two partitions $A$ and $B$,
containing $L_A=(L+1)/2$ and $L_B=(L-1)/2$ vertices, respectively. The
problem Hamiltonian is
\begin{equation}
H_p =
\sum_{i\in A,j\in B}J_{ij}\sigma_i^z\sigma_j^z
+\sum_i\left(\sum_{j\in {\rm nbr}(i)}J_{ij}-2w_i\right)\sigma_i^z,
\label{eq:app_bipartite_Hp}
\end{equation}
where the couplings satisfy
$J_{ij}>\max(w_i,w_j)$. We use the transverse-field driver
\begin{equation}
H_D=-\Gamma\sum_i\sigma_i^x,
\end{equation}
and anneal according to
\begin{equation}
H(s)=sH_p+(1-s)H_D.
\end{equation}

For the uniform-weight construction considered here, let $W_A$ and $W_B$
denote the total weights of the two partitions. When the smaller partition
has the larger total weight, $W_B>W_A$, the two relevant classical
configurations are
\begin{equation}
\ket{A}
=
\ket{\uparrow}^{\otimes L_A}
\ket{\downarrow}^{\otimes L_B},
\qquad
\ket{B}
=
\ket{\downarrow}^{\otimes L_A}
\ket{\uparrow}^{\otimes L_B}.
\end{equation}
Their Hamming distance is $L$. For sufficiently small
$|W_B-W_A|$, these configurations form the two competing low-energy
states at the annealing bottleneck. Since the transverse-field driver
connects them only through $L$ spin flips, their effective coupling arises
at $L$th order in perturbation theory,
\begin{equation}
t_{\rm eff}\sim \Gamma^L
\end{equation}
up to energy-denominator and path-counting factors. Consequently, the
avoided-crossing gap is exponentially small in $L$,
\begin{equation}
\Delta_{\min}\sim e^{-cL}.
\end{equation}

The construction therefore provides a particularly simple example of an
exponentially small gap generated by two low-energy configurations that are
maximally separated in Hamming space. We use this bipartite model in the
main text as the benchmark for the connection between Hamming-space
separation, tunnelling order, and entanglement complexity.

\section{Random graph}
\label{app:appB}

We consider the same random-graph instance used in
Ref.~\cite{hnzj-j7kt}. The graph contains $L=10$ spins, with the
nonzero couplings $J_{ij}$ and local fields $h_i$ specified below. The
construction of the corresponding optimization Hamiltonian is described
in Ref.~\cite{hnzj-j7kt}; we reproduce the numerical parameters here to
fully specify the instance used in the present work.

The local fields are
\begin{equation}
\begin{array}{c|c}
i & h_i\\
\hline
1 & 0.190955\\
2 & 0.823920\\
3 & 0.298059\\
4 & 0.637042\\
5 & 0.799698\\
6 & 0.621208\\
7 & 0.00733097\\
8 & 0.112303\\
9 & 0.553252\\
10 & 0.535868
\end{array}
\end{equation}

The nonzero couplings are
\begin{table}[h]
\centering
\begin{tabular}{c|c@{\qquad}c|c}
$(i,j)$ & $J_{ij}$ & $(i,j)$ & $J_{ij}$\\
\hline
$(1,4)$ & 1.66122 & $(1,6)$ & 1.01834\\
$(1,8)$ & 1.14459 & $(2,3)$ & 1.78942\\
$(2,4)$ & 1.10915 & $(2,6)$ & 1.82820\\
$(2,7)$ & 1.76385 & $(3,4)$ & 1.57587\\
$(3,9)$ & 1.03825 & $(3,10)$ & 1.88831\\
$(4,7)$ & 1.27207 & $(4,9)$ & 1.02395\\
$(4,10)$ & 1.68937 & $(5,6)$ & 1.23293\\
$(5,8)$ & 1.32764 & $(5,9)$ & 1.09610\\
$(6,10)$ & 1.09028 & $(7,8)$ & 1.19425\\
$(7,9)$ & 1.22829 & $(7,10)$ & 1.96842\\
$(8,10)$ & 1.30031 && 
\end{tabular}
\caption{Nonzero couplings defining the random-graph instance used in
this work.}
\label{tab:random_couplings}
\end{table}

The annealing Hamiltonian is
\begin{equation}
H(s)=sH_p+(1-s)H_D,
\qquad
H_D=-\Gamma\sum_{i=1}^{L}\sigma_i^x.
\end{equation}

For this instance, the avoided crossing is dominated by the two
configurations
\begin{equation}
\ket{z_2}=\ket{1100100001},
\qquad
\ket{z_3}=\ket{1010101000},
\end{equation}
which differ by four spin flips,
$d_H(z_2,z_3)=4$. A third configuration,
\begin{equation}
\ket{z_1}=\ket{1100100000},
\end{equation}
also has appreciable weight at the crossing. In particular,
$d_H(z_1,z_2)=1$ and $d_H(z_1,z_3)=3$. The resulting entanglement
structure is discussed in Sec.~\ref{sec:sec2}.
\section{Explicit construction of a logarithmically entangled resonant manifold}
\label{app:appC}

Here we give the details of the construction summarized in Sec.~III\,B.
We consider $L=2n$ data spins divided into equal subsystems $A$ and $B$
($n$ spins each), together with one auxiliary selector spin $C$, for
$L+1$ spins in total. Let
\begin{equation}
x_k=(\underbrace{+\cdots+}_{k}\underbrace{-\cdots-}_{n-k}),
\qquad k=0,\dots,n-1,
\end{equation}
and let $\bar x_k$ denote $x_k$ with every spin flipped. The two classical
manifolds are
\begin{align}
\mathcal G&=\{\ket{x_k}_A\ket{x_k}_B\ket{Z_C{=}{+}1}\}_{k=0}^{n-1},
\nonumber \\
\mathcal E&=\{\ket{\bar x_k}_A\ket{x_k}_B\ket{Z_C{=}{-}1}\}_{k=0}^{n-1}.
\end{align}

\subsection*{Hamming geometry}

Within either manifold,
$d_H(g_k,g_j)=d_H(e_k,e_j)=2|k-j|$, so neighbouring
configurations differ by one spin in $A$ and one in $B$. Between manifolds,
$d_H(x_k,\bar x_j)=n-|k-j|$ on $A$ and
$d_H(x_k,x_j)=|k-j|$ on $B$, so
\begin{equation}
d_H(g_k,e_j)=n=\frac L2
\quad\text{for all }k,j,
\end{equation}
excluding $C$. Because $C$ must also flip, the transverse-field coupling
between the manifolds appears at order
$n+1=L/2+1$.

\subsection*{Problem Hamiltonian}

The Hamiltonian is diagonal and 3-local,
\begin{equation}
H_P=(U_0+U_1Z_C)\,P_B
+\frac K2\sum_{i=1}^{n}\bigl(1-Z^A_iZ^B_iZ_C\bigr)
+\frac\delta2(1-Z_C),
\label{eq:designer_HP}
\end{equation}
with $U_{0,1}=(U_G\pm U_E)/2$ and
\begin{equation}
P_B=\frac{1+Z^B_n}{2}
+\frac14\sum_{i=1}^{n-1}
\bigl(1-Z^B_i+Z^B_{i+1}-Z^B_iZ^B_{i+1}\bigr).
\end{equation}
Since $\tfrac14(1-Z_i)(1+Z_{i+1})$ projects onto a down spin followed by an
up spin, and $\tfrac12(1+Z_n)$ projects onto $Z_n=+1$, $P_B$ vanishes
exactly on $x_0,\dots,x_{n-1}$ and counts violations otherwise. The penalty
is $U_G$ for $Z_C=+1$ and $U_E$ for $Z_C=-1$. The $K$ terms enforce $A=B$
for $Z_C=+1$ and $A=\bar B$ for $Z_C=-1$. For $K,U_G,U_E>0$, $H_P$
therefore has exactly $2n$ states below the penalty scale: the states of
$\mathcal G$ at energy $0$ and those of $\mathcal E$ at energy $\delta>0$.
The annealing Hamiltonian is
\begin{equation}
H(s)=-(1-s)\Gamma\Bigl(\sum_i X^A_i+\sum_i X^B_i+X_C\Bigr)+sH_P.
\label{eq:designer_anneal}
\end{equation}

\subsection*{Perturbative picture and the crossing}

In degenerate perturbation theory in
$(1-s)\Gamma/(sK)\ll1$, each manifold becomes a tight-binding chain.
Moving $k\to k+1$ requires flipping one spin in $A$ and one in $B$; both
intermediate states violate a single three-spin term (energy $sK$), so
\begin{equation}
t\simeq-\frac{2(1-s)^2\Gamma^2}{sK},
\end{equation}
identical in $\mathcal G$ and $\mathcal E$. The resulting low-energy states are delocalized across the respective
manifolds,
\begin{align}
\ket G&=\sum_k a_k\ket{x_k}_A\ket{x_k}_B\ket{Z_C{=}{+}1},
\nonumber \\
\ket E&=\sum_k b_k\ket{\bar x_k}_A\ket{x_k}_B\ket{Z_C{=}{-}1},
\end{align}
with support on $O(n)$ configurations. The two manifolds differ in their diagonal
second-order shifts. Flipping a single $A$ spin costs $K$ in both. Flipping
an interior $B$ spin costs $K+U_C$ and creates one violation of $P_B$,
with $U_C=U_G$ or $U_E$. To leading order the difference in stabilization is
\begin{equation}
\Delta\varepsilon(s)\simeq
n\,\frac{(1-s)^2\Gamma^2}{s}
\left[\frac{1}{K+U_E}-\frac{1}{K+U_G}\right],
\end{equation}
up to edge corrections and higher orders. Hence, within the perturbative
regime, the classically higher manifold $\mathcal E$ is stabilized relative
to $\mathcal G$ when $U_E<U_G$, and the two dressed manifolds cross at
\begin{equation}
s_c\delta\simeq\Delta\varepsilon(s_c).
\end{equation}
Since $\Delta\varepsilon$ is extensive, $\delta\propto n$ keeps $s_c$ fixed;
at fixed $\delta$, $s_c$ drifts toward $1$ with $n$. We work in the regime
\begin{equation}
K,U_{G,E}\gg\frac{(1-s_c)\Gamma}{s_c}.
\end{equation}

\subsection*{Schmidt structure}

The $A$ strings $\{x_k\}\cup\{\bar x_j\}$ are $2n$ distinct configurations,
while the $B$ strings of $\ket G$ and $\ket E$ coincide. Across $A|BC$ the
$2n$ states $(x_k,\pm)$ of $BC$ are distinct, so
$r_S=2n=L$. Across $AC|B$ the two manifolds share $B$ strings and
$r_S=n$. In either case $r_S=\Theta(L)$. Since the low-energy states are
delocalized over $O(n)$ configurations, their Schmidt weights are spread
over an increasing number of branches, giving
\begin{equation}
S=\Theta(\log L).
\end{equation}
Dressing by virtual excitations adds Schmidt components of relative weight
$O(\Gamma_{\rm eff}^2/K^2)$.

\subsection*{Inter-manifold coupling and gap}

Converting $\mathcal G$ into $\mathcal E$ requires one flip per site
(either $A_i$ or $B_i$) and one flip of $C$. Intermediate states can carry
an extensive number of violated three-spin terms, so the energy denominators
grow with the number of flipped spins. Summing over the allowed flip
orderings therefore modifies the prefactor and the exponential rate, but
does not remove the extensive perturbative order. We consequently do not
attempt a sharp prefactor and instead verify the exponential scaling
numerically below. Thus, up to energy-denominator and path-counting factors,
the inter-manifold matrix element has the form
\begin{equation}
t_{\mathcal{GE}}
\sim \left(\frac{\Gamma_{\rm eff}}{K}\right)^n
\end{equation}
with corrections that can modify the coefficient in the exponential.

At $s\to1$, the problem Hamiltonian has an $n$-fold degenerate ground space
consisting of the states in $\mathcal G$, and hence the gap to the next
state within this ground space vanishes at the endpoint. Since every member
of $\mathcal G$ is an optimal solution, this endpoint degeneracy is distinct
from the $\mathcal G$--$\mathcal E$ tunnelling bottleneck of interest here.
We therefore define $\Delta_{\min}$ for this construction as the splitting
at the interior $\mathcal G$--$\mathcal E$ avoided crossing, rather than the
trivial degeneracy at $s=1$.

\end{document}